\documentclass[twocolumn,english,pra,showpacs,final,superscriptaddress]{revtex4}
\usepackage[T1]{fontenc}
\usepackage[latin9]{inputenc}
\usepackage{color}
\usepackage{babel}

\usepackage{amsmath}
\usepackage{graphicx}
\usepackage{amssymb}
\usepackage[unicode=true, pdfusetitle,
 bookmarks=true,bookmarksnumbered=false,bookmarksopen=false,
 breaklinks=false,pdfborder={0 0 1},backref=false,colorlinks=true]
 {hyperref}

\makeatletter
\@ifundefined{textcolor}{}
{%
 \definecolor{BLACK}{gray}{0}
 \definecolor{WHITE}{gray}{1}
 \definecolor{RED}{rgb}{1,0,0}
 \definecolor{GREEN}{rgb}{0,1,0}
 \definecolor{BLUE}{rgb}{0,0,1}
 \definecolor{CYAN}{cmyk}{1,0,0,0}
 \definecolor{MAGENTA}{cmyk}{0,1,0,0}
 \definecolor{YELLOW}{cmyk}{0,0,1,0}
 }

\usepackage{epsfig}

\usepackage{amsfonts}
\usepackage{float}

\makeatother

\usepackage{babel}

\makeatother

\usepackage{babel}

\makeatother

\usepackage{babel}

\makeatother

\begin{document}

\title{Robustness of spin-coupling distributions for perfect quantum state
transfer}

\author{Analia Zwick}

\email{zwick@famaf.unc.edu.ar}

\affiliation{Fakult\"{a}t Physik, Technische Universit\"{a}t Dortmund, D-44221
Dortmund, Germany.}

\affiliation{Facultad de Matem\'{a}tica, Astronom\'{\i}a y F\'{\i}sica and
Instituto de F\'{i}sica Enrique Gaviola, Universidad Nacional de
C\'{o}rdoba, 5000 C\'{o}rdoba, Argentina.}

\author{Gonzalo A. \'{A}lvarez}

\email{galvarez@e3.physik.uni-dortmund.de}

\affiliation{Fakult\"{a}t Physik, Technische Universit\"{a}t Dortmund, D-44221
Dortmund, Germany.}

\author{Joachim Stolze}

\email{joachim.stolze@tu-dortmund.de}

\affiliation{Fakult\"{a}t Physik, Technische Universit\"{a}t Dortmund, D-44221
Dortmund, Germany.}

\author{Omar Osenda}

\email{osenda@famaf.unc.edu.ar}

\affiliation{Facultad de Matem\'{a}tica, Astronom\'{\i}a y F\'{\i}sica and
Instituto de F\'{i}sica Enrique Gaviola, Universidad Nacional de
C\'{o}rdoba, 5000 C\'{o}rdoba, Argentina.}

\keywords{quantum channels, spin dynamics, perfect state transfer, quantum
information, decoherence, mesocopic echoes }

\pacs{03.67.Hk, 03.65.Yz, 75.10.Pq, 75.40.Gb}
\begin{abstract}
The transmission of quantum information between different parts of
a quantum computer is of fundamental importance. Spin chains have
been proposed as quantum channels for transferring information. Different
configurations for the spin couplings were proposed in order to optimize
the transfer. As imperfections in the creation of these specific spin-coupling
distributions can never be completely avoided, it is important to
find out which systems are optimally suited for information transfer
by assessing their robustness against imperfections or disturbances.
We analyze different spin coupling distributions of spin chain channels
designed for perfect quantum state transfer. In particular, we study
the transfer of an initial state from one end of the chain to the
other end. We quantify the robustness of different coupling distributions
against perturbations and we relate it to the properties of the energy
eigenstates and eigenvalues. We find that the localization properties
of the systems play an important role for robust quantum state transfer.
\end{abstract}
\maketitle

\section{\label{sec I:Introduction}Introduction}

Quantum information processing has been extensively studied during
the past years \cite{ladd2010}. One of the main challenges of actual
physical implementations has been the manipulation of the quantum
information with sufficient accuracy to prevent errors. In particular
it is important to be able to transfer quantum information between
different elements of a quantum computer \cite{Divicenzo}. In this
respect, spin chain systems have been proposed as quantum channels
for the transmission of \textit{\emph{quantum states, where the spins
act as the quantum bits}} \cite{Bose2003,Subrahmanyam2004,Linden2004,Li2005,Fitzsimons2006,Roncaglia2007,Plastina2007,Paternostro2008}.
Many systems of this kind have been explored in order to improve their
performance for the state transmission. One of the goals is to find
systems that allow for state transfer without any dynamical manipulations
during the transfer procedure or with only minimal additional requirements.
For example, local control only on the boundary spins in either an
initialized chain \cite{Wojcik2005,Zwick2011} or an unpolarized chain
\cite{Apollaro2010,Lukin2010} can cause a large enhancement of the
transmission fidelity from one end of the chain to the opposite end;
even perfect state transfer (PST) could be achieved by engineering
the entire set of spin-spin couplings in the chain \cite{Albanese2004,Christandl2004,Stolze2005,Kay2006}.

Very few of these systems have been implemented experimentally, for
example using small numbers of spins in liquid state NMR \cite{Madi1997,Nielsen1998,Zhang2005,Suter2007,Alvarez-Frydman2010}
and slightly larger numbers of them in solid-state NMR \cite{Cappellaro2007,Rufeil2009}.
Spin defects in diamond seem to show a promising direction for near
future implementations \cite{Neumann2010,Cappellaro2009,Lukin2010-preprint}.
Important experimental challenges are posed by the lack of individual
addressibility of the spins and, more importantly, by their vulnerability
to decoherence \cite{Zurek2003}. Imperfections in the implementation
of spin-chain systems also cause decoherence and were predicted to
produce localization of the quantum information \cite{Chiara2005,Linden2007,Osborne2007,Linden2009}
which was recently demonstrated experimentally \cite{AlvarezPRL2010}.
Consequently, a successful characterization of these PST protocols
should consider these errors in order to find the optimal one. Two
PST protocols that require engineered spin-couplings \cite{Christandl2004,Stolze2005}
have been analyzed in this respect considering static perturbations
\cite{Chiara2005,Damico2011,Kay2006}. Other important points to consider
are the timing errors on the time when the PST is achieved \cite{Kay2006}
and the speed of transfer of the different protocols \cite{Young2006}.
But, considering that the number of possible systems that could be
used for PST \cite{Stolze2005,Kay2006} is infinite, a performance
comparison between them should be aimed at finding a system which
is as robust against perturbations as possible. For that purpose it
is important to find out which intrinsic properties of a system make
it robust against perturbations. In this work, we tackle these questions
analyzing different energy distributions that allow for PST and compare
their robustness against static perturbations. We characterize the
robustness of the systems by calculating their transmission fidelity.
In order to find the relevant properties of the systems that make
them robust, we analyze how the eigenstates and eigenenergies are
perturbed. We find that the localization properties of a system are
intimately connected to its robustness.

The paper is organized as follows, in Sec. \ref{sec II:Perfect-state-transfer-channels}
we present the XX model describing the quantum spin chain and the
necessary conditions for perfect state transfer. In Sec. \ref{sec III:Energy-and-spin-coupling}
we analyze different energy eigenvalue configurations of the system
and the corresponding spin-coupling distributions. In Sec. \ref{sec IV - sub A: Non-perturbed-transfer}
we analyze the fidelity of the transfer of the different configurations,
and the influence of perturbations on the transmission is discussed
in Section \ref{sec IV - sub B: :Perturbed-transfer}. Subsequently,
in Sec. \ref{sec V:Energy-levels-contribution}, we analyse how the
individual perturbed eigenstates and eigenvalues contribute to the
dynamics of quantum information transport. Finally, in Sec. \ref{sec VI:Conclusion}
we give the conclusions.

\section{\label{sec II:Perfect-state-transfer-channels}Perfect state transfer
channels}

We consider a chain of $N$ spins 1/2 (qubits) with a modulated XX
interaction between nearest neighbors. Taking into account an external
magnetic field, the Hamiltonian is \begin{equation}
H=\sum_{i}\frac{J_{i}}{2}(\sigma_{i}^{x}\sigma_{i+1}^{x}+\sigma_{i}^{y}\sigma_{i+1}^{y})-\sum_{i}\frac{b_{i}}{2}\sigma_{i}^{z},\label{hamiltonian}\end{equation}
 where $\sigma_{i}^{u}$ are the Pauli operators of the $i$th spin,
$b_{i}$ is the local external field and $J_{i}$ is the exchange
coupling.

The aim is to transmit a quantum state $\left|\psi_{0}\right\rangle $
stored on the first spin ($i=1$) to the last spin of the chain ($i=N$),
where $\left|\psi_{0}\right\rangle =\alpha\left|0\right\rangle +\beta\left|1\right\rangle $
is a given superposition of a spin down and up respectively and the
remaining spins of the chain are initialized in the spin down state.
The Hamiltonian (\ref{hamiltonian}) preserves the total magnetization
along the $z$-axis because $[H,\Sigma_{i}\sigma_{i}^{z}]=0$, \emph{i.e.},
the number of excited spins is conserved. Because the initial state
$|\Psi_{0}\rangle$ is a superposition of the eigenstate $|\mathbf{0}\rangle=|00...0\rangle$
and the state $\mathbf{|1}\rangle=\mathbf{|}1_{1}0......0\rangle$,
the component $|\mathbf{0}\rangle$ is conserved and the component
$\mathbf{|1}\rangle$ evolves within the one excitation subspace spanned
by the basis states $\mathbf{|i}\rangle=\mathbf{|}0...01_{i}0...0\rangle$.
The state of the system at a given evolution time $t$ is\begin{equation}
|\Psi(t)\rangle=e^{-iHt/\hslash}|\Psi_{0}\rangle=\alpha|\mathbf{0}\rangle+\beta\sum_{i=1}^{N}f_{i}(t)|\mathbf{i}\rangle,\end{equation}
 where $f_{i}(t)=\langle\mathbf{i}|e^{-iHt/\hslash}|\mathbf{1}\rangle$.
To measure the effectiveness of state transfer between sites 1 and
$N$, we determine the fidelity $\mathcal{F}(t)=\langle\Psi_{0}|\rho_{N}(t)|\Psi_{0}\rangle$
averaged over all possible initial states $|\Psi_{0}\rangle$ distributed
uniformly over the Bloch sphere, which is given by \cite{Bose2003}
\begin{equation}
F(t)=\frac{|f_{N}(t)|\cos\gamma}{3}+\frac{|f_{N}(t)|^{2}}{6}+\frac{1}{2},\label{averagefidelity}\end{equation}
 where $\gamma=\arg{|f_{N}(t)|}$. Because the phase $\gamma$ can
be controlled by an external field once the state is transferred,
we consider $\cos\gamma=1$. PST is achieved when $F=1$.

For a spin chain possessing mirror symmetry with respect to the center,
\textit{i.e.,} $J_{i}^{2}=J_{N-i}^{2}$ and $b_{i}=b_{N+1-i}$, the
necessary and sufficient condition for PST is \begin{equation}
\omega_{k+1}-\omega_{k}=(2m_{k}+1)\pi/t_{\mathrm{PST}},\label{spectrum}\end{equation}
where the set of eigenenergies $\{\omega_{k}\}$ is ordered, $\omega_{k}<\omega_{k+1}$.
The condition (\ref{spectrum}) must be fulfilled for all pairs of
successive energies, where the $m_{k}$ may be arbitrary integers.
The shortest time $t_{\mathrm{PST}}$ for which (\ref{spectrum})
is fulfilled is the first time at which PST is achieved \cite{Stolze2005,Kay2010}.
Since (\ref{spectrum}) implies strictly periodic time evolution,
PST occurs again and again, at all odd multiples of $t_{\mathrm{PST}}$.

\section{\label{sec III:Energy-and-spin-coupling}Energy and spin-coupling
distributions}

Every set of integers $m_{k}$ in (\ref{spectrum}) leads to a unique
energy spectrum enabling PST and hence, as we shall explain below,
to a unique set of coupling constants $J_{i}$. Therefore, there are
infinitely many spin chains allowing for PST. But, are all of them
equally efficient for transferring information? How is their PST capability
affected by perturbations through inaccuracies in the coupling constants
or from coupling to external degrees of freedom? What properties are
necessary to stabilize the system against such perturbations?

We tackle these questions by studying the transmission robustness
of different PST channels in the presence of static perturbations.
We characterize these spin-channel systems by their energy eigenvalue
distributions. A given spectrum that satisfies the condition (\ref{spectrum})
defines a unique Hamiltonian with positive symmetric couplings $J_{i}$,
which can be obtained by solving an inverse eigenvalue problem \cite{IEP}.
For simplicity we choose $\omega_{k}=-\omega_{N+1-k}$, $\forall k$,
which imposes $b_{i}=0$, $\forall i$ \cite{Kay2010}.

\begin{figure}[h]
\begin{centering}
\includegraphics[width=8.5cm]{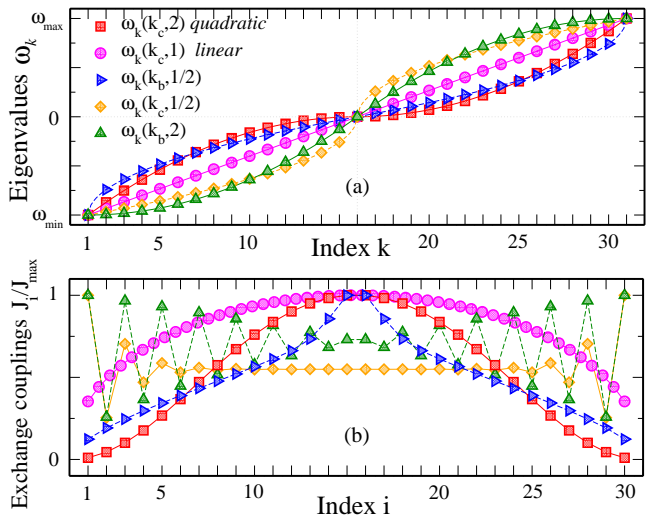} 
\par\end{centering}

\caption{\label{fig01}(Color online) (a) Energy eigenvalue distributions $\omega_{k}(k_{\beta},\alpha)$.
The symbols represent the energy values and the lines give the exact
functional dependence of Eq. (\ref{omega}). (b) Exchange couplings
determined by solving the inverse eigenvalue problem for each of the
spectra given in (a).}

\end{figure}

In order to study a range of different eigenvalue distributions systematically,
we start from the case of an equidistant energy spectrum, $m_{k}=$
const. in (\ref{spectrum}), which was discussed in Ref. \cite{Christandl2004}.
We vary that spectrum by distributing the energy values more densely
either in the center or towards the boundaries of the energy spectrum.
The class of energy spectra which we discuss can be parametrized as
follows: \begin{equation}
\omega_{k}(k_{\beta},\alpha)=-\mbox{\ensuremath{A_{(k_{\beta},\alpha)}}}\mathrm{sgn}(k-k_{0})[(k_{\beta}-|k-k_{0}|)^{\alpha}-k_{\beta}^{\alpha}].\label{omega}\end{equation}
We assume that $N$ is odd, $k=1,...,N$ numbers the energy eigenvalues
in ascending order, as before, and $k_{0}=\frac{N-1}{2}$ marks the
center of the spectrum. The shape of the spectrum is controlled by
an exponent $\alpha$ and a reference index $k_{\beta}$ which can
assume two values; $k_{\beta}=k_{b}=k_{0}$ or $k_{\beta}=k_{c}=0$.
The overall width of the spectrum is controlled by $A_{(k_{\beta},\alpha)}$.
The equidistant energy spectrum (constant density of eigenvalues)
is given by $\omega_{k}(k_{c},1)$. The density of eigenvalues in
the center of the spectrum increases for both $\omega_{k}(k_{c},n)$
and $\omega_{k}(k_{b},\frac{1}{n})$ with integer $n\geq2$. A larger
density of eigenvalues close to the boundaries of the spectrum is
obtained for $\omega_{k}(k_{b},n)$ and $\omega_{k}(k_{c},\frac{1}{n})$.
The shapes of the two spectra defined by these two possibilities for
given $n$ are different, as are those of $\omega_{k}(k_{c},n)$ and
$\omega_{k}(k_{b},\frac{1}{n})$, respectively. For non-integer exponent
$\alpha$ the energies (\ref{omega}) normally do not fulfill the
commensurability condition (\ref{spectrum}) and have to be slightly
readjusted to make PST possible. Figure \ref{fig01}(a) shows the
energy eigenvalues for the equidistant spectrum, $\omega_{k}(k_{c},1)$,
along with the four possibilities just discussed, for $n=2$. The
corresponding exchange couplings $J_{i}$ (normalized by the maximum
coupling strength $J_{max}$) are shown in Fig. \ref{fig01}(b). The
coupling distribution determines the transmission velocity as we shall
discuss in Sec. \ref{sec IV:Perfect-state-transfer-stability-of-energy-distributions}.

\section{\label{sec IV:Perfect-state-transfer-stability-of-energy-distributions}Perfect
state transfer stability of energy distributions}

\subsection{\label{sec IV - sub A: Non-perturbed-transfer}Unperturbed transfer}

\begin{figure}[h]
\begin{centering}
\includegraphics[width=8.6cm]{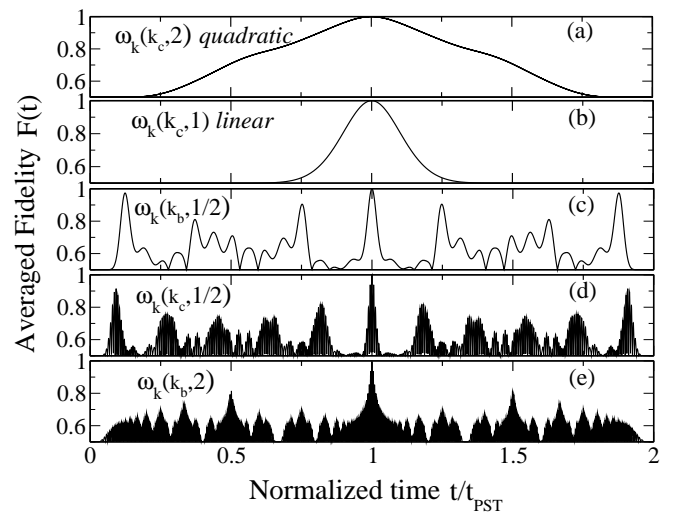} 
\par\end{centering}

\caption{\label{fig02}Averaged fidelity of the state transfer in a $N=31$
spin chain for the different energy distributions shown in Fig. \ref{fig01}
as a function of time. The \textit{linear} $\omega_{k}(k_{c},1)$
and \textit{quadratic} $\omega_{k}(k_{c},2)$ distributions achieve
the perfect transmission with the first echo, while the other cases
achieve it after several echoes. In panels (d) and (e), the black
regions are due to fast oscillations because of the coupling strength
oscillations shown in Fig. \ref{fig01}b. The transfer times are given
by $t_{PST}^{linear}=\frac{\pi N}{4J_{max}}$ and $t_{PST}\sim\gamma t_{PST}^{linear}$
with $\gamma=15.4,\,17,\,15,\,14.5$ for a $quadratic,\,\omega_{k}(k_{b},\frac{1}{2}),\,\omega_{k}(k_{c},\frac{1}{2})$
and $\omega_{k}(k_{b},2)$\textbf{ }distribution respectively.}
 
\end{figure}

To compare the perfect state transfer performance of the spin-channels
with the different energy eigenvalue distributions of Fig. \ref{fig01},
we calculated their averaged fidelity (\ref{averagefidelity}). Figure
\ref{fig02} shows the fidelity of state transfer from one end of
the chain to the other, as a function of time. The time scale is given
in units of the first perfect state transfer time $t_{\mathrm{PST}}$.
At this point it is important to note that the dynamics of the system
contain at least two other relevant time scales besides the time $t_{\mathrm{PST}}$
which we shall use as a unit of time. The first such time scale is
the spin-channel clock time $2t_{M}$, \emph{i.e.}, the characteristic
time of the information propagation within the chain, also called
mesoscopic echo time \cite{Altshuler94}. For a chain supporting spin
waves as elementary excitations, e.g. the uniformly coupled XX or
Heisenberg chains, $t_{M}$ is fixed by the maximum group velocity
of the spin waves \cite{Pastwaski95,Pastawski96,Feldman1998,Bose2003}.
The group velocity of excitations with dispersion $\omega(k)$ (where
$k$ now temporarily denotes the wave number) is given by $v_{g}=\frac{d\omega}{dk}$.
Unfortunately this concept breaks down for the systems of interest
here, since translational invariance is broken by the non-uniform
couplings $J_{i}$ and the wave number is no longer defined. Our numerical
results show that $t_{PST}$ can be larger than $t_{M}$, see for
example, Fig. \ref{fig02}(c) and (d). At $t_{M}$ the excitations
created at $t=0$ at site $i=1$ interfere constructively but not
perfectly at site $i=N$. Perfect interference occurs only later,
at $t=t_{PST}$, after the excitations have travelled back and forth
between the ends of the chain many more times. Fig. \ref{fig02}(a)
and (b) show that the \textit{linear} $\omega_{k}(k_{c},1)$ and \textit{quadratic}
$\omega_{k}(k_{c},2)$ distributions achieve perfect transfer without
secondary maximum of the fidelity at some earlier time. For those
two systems $t_{PST}$ is thus equal to the $t_{M}$ of the spin chain.
To make a quantitative analysis of the speed of the transfer, we consider
as a reference the known value $t_{M}^{h}$ for a homogeneous spin-chain
with $J_{i}\equiv J$ in Eq. (\ref{hamiltonian}). In that system,
constructive interference at site $N$ occurs at time $t_{M}^{h}\sim\frac{N}{2J}$
\cite{Feldman1998}. The transfer obtained at that instant is not
perfect, but by switching couplings on and off to perform consecutive
swap operations, perfect transfer may be achieved at $t_{M}^{swap}\sim\frac{\pi N}{2J}$
\cite{Petrosyan2010}. In terms of the maximum coupling $J_{max}$,
the PST time for the linear distribution is $t_{PST}^{linear}=\frac{\pi N}{4J_{max}}$,
which is two times faster than the consecutive swaps assuming $J=J_{max}$,
$t_{PST}^{linear}=\frac{1}{2}t_{M}^{swap}$, but slower than the free
evolution, $t_{PST}^{linear}=\frac{\pi}{2}t_{M}^{h}$, in a homogeneous
chain. The other distributions are about 15 times slower than the
linear case as listed in the caption of \ref{fig02}. The second important
time scale is given by the duration of the PST maximum of the fidelity,
i.e. the time during which the fidelity is very close to unity. $\Delta t$
can be interpreted as the time of residence of the perfectly transmitted
state on the last site of the chain; it determines the timing precision
required for perfect state read-out. While the \textit{quadratic}
distribution is much slower than the linear one in terms of transfer
time, its advantage is a much longer window time. We will return to
this point later in Sec. \ref{sec V:Energy-levels-contribution}.

\subsection{\label{sec IV - sub B: :Perturbed-transfer}Perturbed transfer}

So far, we have discussed the performances of different spin-channels
without any external perturbation. However, since the perfect engineering
of all spin couplings is highly improbable, the study of the performance
of different spin-coupling distributions under perturbations by flawed
spin couplings becomes relevant. To study the robustness of the spin
chains against perturbations we introduce static random spin-coupling
imperfections quantified by $\delta_{i}$ \begin{equation}
J_{i}\longrightarrow J_{i}(1+\delta_{i}),\end{equation}
 where each $\delta_{i}$ is an independent uniformly distributed
random variable in the interval $[-\varepsilon_{J},\varepsilon_{J}]$.
$\varepsilon_{J}$ is a positive real number that characterizes the
maximum perturbation strength relative to $J_{i}$.

\begin{figure}[h]
 \includegraphics[width=8.7cm]{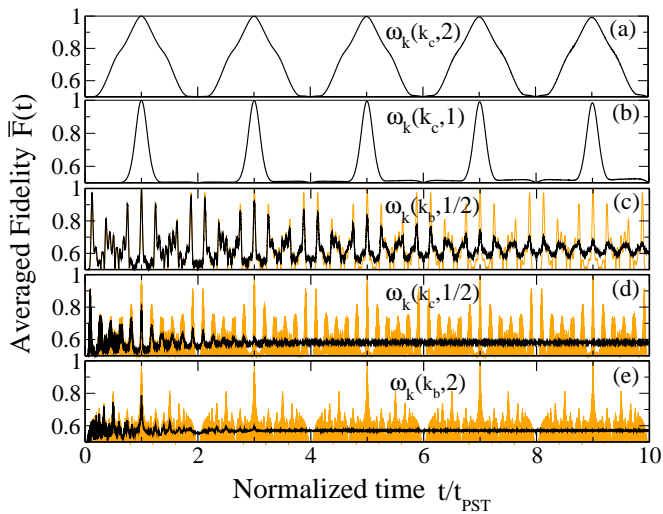} \caption{\label{fig03} (Color online) Averaged fidelity of the state transfer
in a $N=31$ spin chain with random perturbations of strength $\varepsilon_{J}=10^{-2}$
averaged over $N_{av}=10^{2}$ realizations for the different energy
distributions from Fig. \ref{fig01} as a function of time. The colored
lines in (c), (d) and (d) show $F(t)$ for the systems without perturbation
($\varepsilon_{J}=0$).}

\end{figure}

We calculate numerically the fidelity time evolution $\overline{F}(t)={\langle{F}(t)\rangle}_{N_{\mathrm{av}}}$
averaged over $N_{\mathrm{av}}$ different realizations of the random
imperfection values $\delta_{i}$. Figure \ref{fig03} shows the averaged
fidelity evolution for the different energy eigenvalue distributions
for a common $\varepsilon_{J}$ value. Only two cases are strongly
distinguished by their robustness against the perturbation: the \textit{linear}
distribution, which was already studied by De Chiara et. al. \cite{Chiara2005},
and the \textit{quadratic} distribution. As the near-perfect echoes
in Fig. \ref{fig03}(a) and (b) show, disorder at the level of $\varepsilon_{J}=10^{-2}$
does not significantly affect PST in those two cases. In contrast,
panels (c) and especially (d) and (e) of the same figure show a rather
rapid decay of the fidelity (black line) down to a useless level.
The colored lines in panels (c), (d) and (e) show the fidelities of
the unperturbed systems for comparison. %

\begin{figure}[h]
 \centering{}\includegraphics[width=8cm]{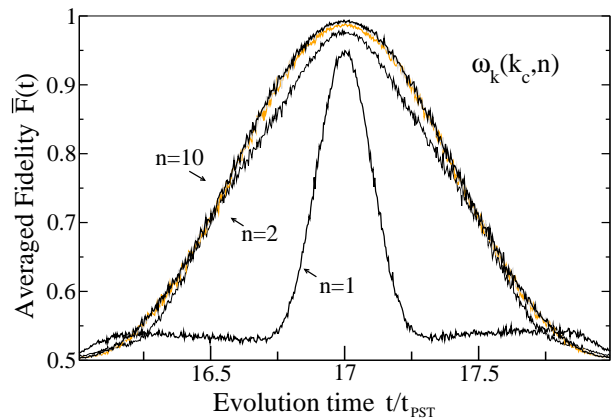} \caption{\label{fig04} (Color online) Averaged fidelity of the state transfer
in a $N=31$ spin chain with random perturbations of strength $\varepsilon_{J}=10^{-2}$
averaged over $N_{av}=10^{2}$ realizations for eigenvalue distributions
$\omega_{k}(k_{c},n)$. Shown is the range of times around the 9th
echo of the PST in the unperturbed chain, for $n=1,2,10$ (black lines),
and $n=3$ (faint colored line very close to the $n=10$ results).}

\end{figure}

Figure \ref{fig04} shows in detail the comparison of the 9th echo
between the\textit{ linear} and the \textit{quadratic} distributions.
The \textit{quadratic} distribution is obviously more robust than
the \textit{linear} distribution, and also its $\Delta t$ is larger.
Also shown are the 9th echoes for $n=3$ (colored line) and for $n=10$,
respectively. These data show that both the maximum fidelity and the
length $\Delta t$ of the time window for the state read-out increase
with $n$ for energy eigenvalue distributions of type $\omega_{k}(k_{c},n)$.
However, the increase from $n=3$ to $n=10$ is insignificant compared
to the increase from $n=2$ to $n=3$.%

The increase of $\Delta t$ with growing $n$ can be explained by
the changes of the exchange couplings $J_{i}$ shown in Fig. \ref{fig01}(b).
When $n$ changes from 1 to 2 the $J_{i}$ decrease close to the boundaries
and increase in the center of the chain. This trend continues even
more strongly for larger values of $n$ (data not shown). The small
spin couplings close to the boundaries of the chain prevent the spreading
of the information once it is localized at one of the chain ends,
thus leading to larger values of $\Delta t$.

\begin{figure}[h]
\centering{}\includegraphics[width=8cm]{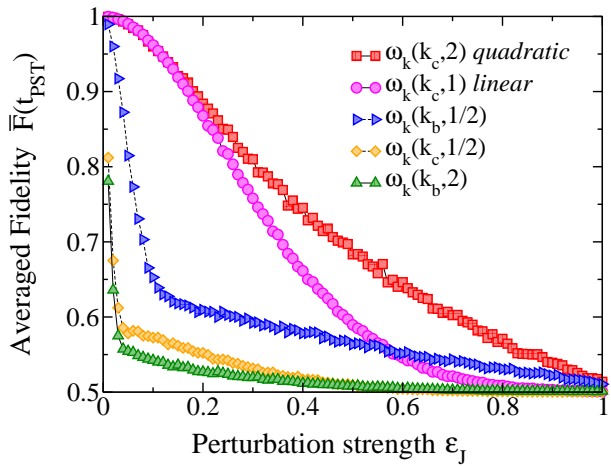} \caption{\label{fig05} (Color online) Averaged fidelity at time $t_{PST}$
as a function of the perturbation strength $\varepsilon_{J}$ for
the different energy distributions from Fig. \ref{fig01} with $N=31$
and $N_{av}=10^{2}$.}

\end{figure}

We now focus on $n=2$ because the robustness properties are similar
for larger values of $n$, but it should be kept in mind that energy
eigenvalue distributions $\omega_{k}(k_{c},n)$ with larger $n$ are
generally more robust. To determine the robustness of the different
distributions, we calculate the averaged fidelity $\overline{F}(t_{\mathrm{PST}},\varepsilon_{J})$
as a function of the perturbation strength $\varepsilon_{J}$ for
the first PST time $t_{\mathrm{PST}}$ determined from the unperturbed
case. Figure \ref{fig05} shows results for different energy distributions
and for a wide range of perturbation strengths. The \textit{linear}
and \textit{quadratic} distributions turn out to be the most robust
ones for all perturbation strengths of interest, yielding quite similar
results for weak perturbations ($\varepsilon_{J}\lesssim0.2$) where
the fidelity is larger than $F=0.9$. For larger perturbation strengths,
the \textit{quadratic} distribution is most robust.

\begin{figure}[h]
 \includegraphics[width=8.5cm]{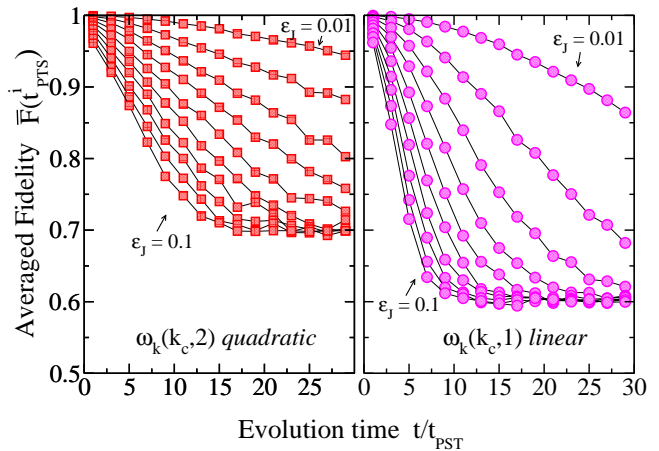} \caption{\label{fig06} (Color online) Averaged fidelity at odd multiples of
$t_{PST}$ (symbols) for the linear and quadratic energy eigenvalue
distributions, $\omega_{k}(k_{c},n)$ $(n=1,2)$. Chain length is
$N=31$, averages were performed over $N_{av}=10^{2}$ realizations.
Perturbation strengths are $\varepsilon=0.01,$ 0.02, ..., 0.1.}

\end{figure}

Recently, it has been shown that the relative decay of the mesoscopic
echoes between a perturbed evolution and the corresponding unperturbed
evolution could be used to determine and characterize the decoherence
time of the spin-chain channel \cite{Alvarez2010}. Similarly, to
determine the decoherence time for each perturbation strength, we
study the state transfer fidelity for different PST echoes as a function
of their respective PST echo times $t_{\mathrm{PST}}^{i}=(2i-1)t_{\mathrm{PST}}$,
\emph{i.e.}, the times where the \textit{i-th} PST echo arrives at
site $N$ for an unperturbed evolution. Figure \ref{fig06} shows
the fidelity $\overline{F}(t_{\mathrm{PST}}^{i},\varepsilon_{J})$
as a function of $t_{\mathrm{PST}}^{i}$, for different perturbation
strengths $\varepsilon_{J}$. The left panel shows the fidelity for
the \textit{quadratic} distribution while the right panel illustrates
the \textit{linear} distribution. The\emph{ }\textit{\emph{decoherence
time, }}\textit{i.e.}\textit{\emph{, the decay time as a function
of }}$t_{\mathrm{PST}}^{i}$\textit{\emph{ is longer for the}} \textit{quadratic}
distribution than for the linear one. Additionally the fidelity of
the quadratic distribution converges to an asymptotic value higher
than that of the \textit{linear} distribution due to the localization
effects caused by the small couplings in the borders, as discussed
above. Weak coupling between terminal qubits and the intervening spin
chain were used as key elements also in other proposals for quantum
information transfer by spin chains recently \cite{Wojcik2005,Zwick2011,Apollaro2010,Lukin2010,Oh2011}.

\section{\label{sec V:Energy-levels-contribution}Robustness and Localization}

\begin{figure*}[ht]
 \includegraphics[clip,width=1\textwidth]{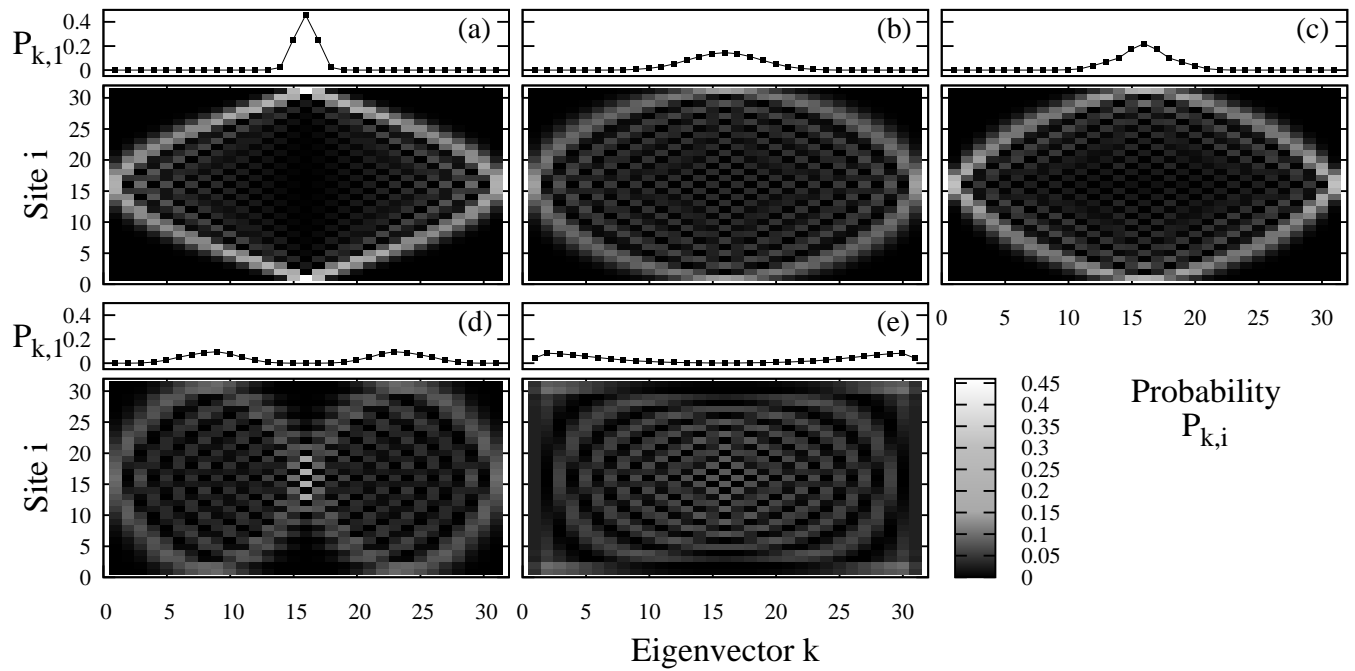} \caption{\label{fig07} Eigenvector probability $P_{k,i}$ of the site (computational)
states $|i\rangle$. $P_{k,i}=a_{k,i}^{2}$, where $|i\rangle=\sum_{k}a_{k,i}|\Psi_{k}\rangle$.
The top part of each panel shows the probabilities $P_{k,1}$ of the
initial state $|\Psi_{0}\rangle=|1\rangle$, and thus shows which
energy eigenstates contribute to the state transfer. The panel labels
refer to the different energy distributions given in Fig. \ref{fig01},
where a) $\omega_{k}(k_{c},2)$ \textit{quadratic}, b) $\omega_{k}(k_{c},1)$
\textit{linear} c) $\omega_{k}(k_{b},\frac{1}{2})$, d) $\omega_{k}(k_{c},\frac{1}{2})$,
and e) $\omega_{k}(k_{b},2)$.}

\end{figure*}
We have shown that certain systems are more robust against perturbations
than others. In order to optimize the engineered spin coupling distributions
it is decisive to understand which properties of the system are relevant
for the robustness of the perfect state transfer. To this end we determine
how each energy eigenstate contributes to the dynamics for each of
the spin-channels. We expand the states $|i\rangle$ (a single excitation
at site $i$) in the eigenstate basis $|i\rangle=\sum_{k}a_{k,i}|\Psi_{k}\rangle$,
where $k$ numbers the energy eigenstates in ascending order, as usual.
Figure \ref{fig07} shows the weights $P_{k,i}=|a_{k,i}|^{2}$, for
the different energy spectra from Fig. \ref{fig01}. The mirror symmetries
with respect to both the center of the chain and the center of the
energy spectrum are due to the spatial mirror symmetry of the couplings,
and the symmetric tridiagonal nature (with zero diagonal) of the Hamiltonian
matrix, respectively. Under perturbations of the couplings the spatial
symmetry of the patterns of Fig. \ref{fig07} is destroyed, while
the energetic symmetry is not. We observe that the degree of localization
of the energy eigenstates varies strongly between the different eigenvalue
distributions. The most robust distributions seem to generate the
most strongly localized energy eigenstates; in panels (a), (b) and
(c) of Fig. \ref{fig07} each energy eigenstate basically seems to
be localized on two lattice sites. The quadratic distribution {[}panel
(a){]} seems to have the most strongly localized eigenstates; in particular
the eigenstates that belong to the center of the band are highly localized
on the boundaries. This is particularly clear from the upper panels
in Fig. \ref{fig07}, showing the contributions $P_{k,1}$ of the
energy eigenstates $|k\rangle$ to the initial state $|i=1\rangle$
with a single excitation localized at the boundary of the chain. In
comparison, the other energy distributions show a larger spread in
the contributions of the energy eigenstates to each site eigenstate
$|i\rangle$. Nevertheless, we observe similarities of the distribution
of $P_{k,i}$ between the \textit{linear,} \textit{quadratic}\textit{\emph{
and}}\textit{ }$\omega_{k}(k_{b},\frac{1}{2})$ distributions. It
has been shown that the presence of localized states at the boundaries
of the spin chain can improve the transmission of quantum states \cite{Zwick2011,Lukin2010,Gualdi2008,Stolze2011}.
These localized states arise when the coupling of the boundary sites
is weaker than the coupling between inner sites or if external fields
are applied at the boundary sites. Therefore, we study how the different
energy levels are affected by perturbations for the different energy
distributions. 

\begin{figure}[h]
 \includegraphics[width=8cm]{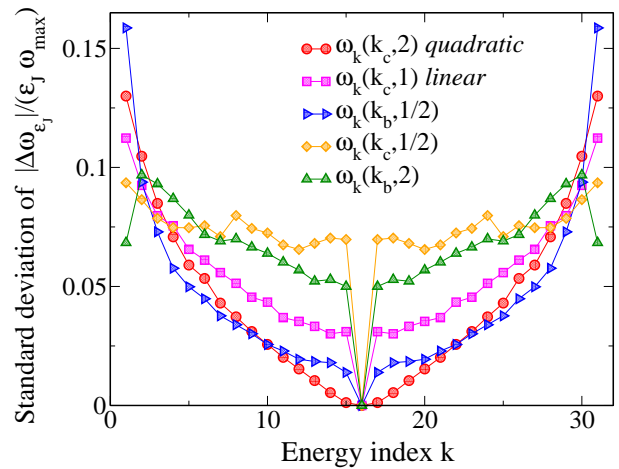} \caption{\label{fig08}(Color online) Standard deviation of the energy levels
$|\Delta\omega_{\varepsilon_{J}}|=|\omega_{\varepsilon_{J}}-\omega_{0}|$
due to the perturbation with strength $\varepsilon_{J}$ for the different
energy distributions of Fig. \ref{fig01}. For weak perturbations
(small $\varepsilon_{J}$) the standard deviation turns out to be
proportional to $\varepsilon_{J}\omega_{max}$, which we use as a
unit here. The data shown are for $\varepsilon_{J}<0.1$; $N_{av}=10^{3}$
realizations were used for the calculations.}

\end{figure}

We generated distributions of energy eigenvalues $\omega_{k}$ for
the different kinds of unperturbed energy spectra and for different
perturbation strengths $\varepsilon_{J}$. For small $\varepsilon_{J}$
we observe a symmetric distribution of the perturbed eigenvalues $\omega_{k}$
around their respective unperturbed values. The width of that distribution
scales with the perturbation strength. For larger values of $\varepsilon_{J}$
the distributions of the perturbed $\omega_{k}$ become asymmetric
with respect to the unperturbed energy level; the low-lying levels
tend to be pulled down, while the high-lying levels are pushed up
by the same amount. (The energy spectrum of the perturbed Hamiltonian
matrix is still symmetric.) The value of $\varepsilon_{J}$ where
the asymmetry sets in depends on the type of unperturbed energy spectrum
and is largest for the quadratic case. To see more quantitatively
what is going on in detail, we show in Fig. \ref{fig08} the standard
deviations of the energy levels for the different kinds of unperturbed
spectra. Each data point represents an average over $N_{av}=10^{3}$
realizations of the random perturbations.%
 The symmetry of the data with respect to the center $\omega=0$ of
the energy spectrum and the fact that the zero energy eigenvalue is
not affected by the randomness at all are due to the nature (symmetric,
tridiagonal, zero diagonal elements) of the Hamiltonian matrix. The
key observation explaining the differences in state transfer robustness
is made by combining figures \ref{fig08} and the upper panels of
Fig. \ref{fig07}. Those panels show that for the quadratic energy
spectrum $\omega_{k}(k_{c},2)$ the initial state $|i=1\rangle$ of
the state transfer process is superposed from a small number of energy
eigenstates in the center of the energy spectrum. In all other types
of energy spectrum the initial state shows a wider distribution in
the energy quantum number $k$. At the same time, the sensitivity
to perturbations (which is what is shown in Fig. \ref{fig08}) shows
a comparatively wide minimum with value zero in the center of the
spectrum, whereas all other types of spectra show roughly constant
nonzero values in the central region of the spectrum, with a single
exceptional zero right in the center. The observed particular robustness
of the \textit{quadratic} energy spectrum can thus be ascribed to
the fact that the initial state consists of a particularly small number
of energy eigenstates coming from a part of the energy spectrum which
is particularly insensitive to perturbations in the spin coupling
constants. For the less robust distributions, the variance is roughly
independent of the energy in the center of the energy band, while
for the quadratic energy spectrum the variance decreases continuously
towards the band center. For that distribution (and for the two other
distributions shown in the upper panels of Fig. \ref{fig07}) all
energy eigenstates are quite strongly localized. A glance at Fig.
\ref{fig01}b shows that the corresponding coupling patterns have
the smallest couplings close to the ends of the chain, in the region
where those energy eigenstates are localized which are most important
for the state transfer. Since we discuss a constant \emph{relative}
strength $\varepsilon$ of the disorder, the \emph{absolute} changes
of the couplings tend to be smallest near the ends of the chain, causing
only small changes in the energy eigenvalues. That explains the particular
robustness of the quadratic distribution. 

Another important aspect characterizing the robustness of the transmission
is the length of the window of time where high fidelity is obtained
for the transmitted state. In this context, we are not only considering
errors in the engineered spin couplings, but also the timing error
of the measurement \cite{Kay2006}. To that end we analyze the term
$|f_{N}(t)|^{2}=|\langle\mathbf{N}|e^{-iHt/\hslash}|\mathbf{1}\rangle|^{2}$
from (\ref{averagefidelity}) at time $t_{PST}+\delta t$. Taking
into account the spatial symmetry, $|f_{N}(t_{PST}+\delta t)|^{2}$
is given by

\begin{align}
|f_{N}|^{2}= & |\sum_{k,s}(-1)^{k+s}P_{s,1}P_{k,1}e^{-i(\omega_{k}-\omega_{s})(t_{PST}+\delta t)}|\nonumber \\
\approx & |\sum_{k,s}P_{s,1}P_{k,1}(1-\delta t(\omega_{k}-\omega_{s})+\nonumber \\
 & i^{2}\frac{\delta t^{2}}{2!}(\omega_{k}-\omega_{s})^{2}-...)|\nonumber \\
\approx & 1-\frac{\delta t^{2}}{2!}\sum_{k,s}P_{s,1}P_{k,1}(\omega_{k}-\omega_{s})^{2},\end{align}
 where $P_{k,i}=|\langle\Psi_{k}|1\rangle|^{2}$. Even without the
Bloch-sphere average (\ref{averagefidelity}) which would be necessary
for a comparison with Fig. \ref{fig02}, the above result shows why
the quadratic distribution displays the longest window of time. The
probabilities $P_{k,1}$ (see Fig. \ref{fig07}) are sharply peaked
in the center of the energy band and essentially zero otherwise. Furthermore,
due to the quadratic nature of the energy spectrum, the relevant energy
differences $\omega_{k}-\omega_{s}$ are particularly small (see Fig.
\ref{fig01}a) making the fidelity deviate from unity only at rather
large $\delta t$ values.\\

\section{\label{sec VI:Conclusion}Conclusion}

We have studied the robustness of spin chain systems designed for
perfect quantum state transfer (PST) under static perturbations. We
explored different PST systems by choosing different energy spectra
distributions that satisfy the PST conditions. From the energy spectrum
of a given chain, the spin-spin coupling constant pattern can be obtained
by solving an inverse eigenvalue problem. The robustness of each system
was studied by calculating its transmission fidelity under static
perturbations of the couplings. We found that robustness is characterized
by two main features. One is the reduction of the transfer fidelity
induced by the perturbed couplings and the other is the duration of
the time window during which the transmitted state may be read out
with high fidelity. The most robust systems are those with linear
and quadratic energy eigenvalue distributions. These systems achieve
PST at the time of the first fidelity maximum. That time may be called
the spin-wave echo time, and the less robust systems reach PST only
after several spin-wave echoes. By analyzing how the energy eigenstates
and eigenvalues are affected by the perturbations, we found that the
most robust distributions have strongly spatially localized eigenstates.
Thus, because the initial state is localized in one end of the chain,
only few eigenstates participate in the transfer. Because of the localization
properties of the eigenstates the perturbations in the spin couplings
close to the chain boundaries are the only significant source of errors.
Since these couplings are rather weak for the most robust systems,
a given relative perturbation strength only causes a small absolute
perturbation in the couplings and thus in the energy eigenvalues,
leading to the observed robustness. The weak couplings close to the
ends of the chain also lead to a longer residence time of the transmitted
state at its target site at the chain boundary, causing a longer time
window for read-out. 

\begin{acknowledgments}
We acknowledge support from SECYT-UNC (Universidad Nacional de C\'ordoba,
Argentina), and CONICET (Argentina) for partial financial support
of this project. A. Z. thanks CONICET and DAAD, and G. A. A. the Alexander
von Humboldt Foundation, for their Research Scientist Fellowships,
and the hospitality of Fakult\"{a}t Physik of TU Dortmund. We also
thank P. Karbach for helpful discussions and for his inverse eigenvalue
program.\end{acknowledgments}

\end{document}